\documentclass[aps,prl,twocolumn,showpacs,psfig,superscriptaddress]{revtex4-1}
\usepackage{times}
\usepackage{graphicx}
\usepackage{float}
\usepackage{latexsym,amsmath,amssymb,bm,euscript}
\usepackage{color}
\usepackage{subfigure}
\usepackage{epstopdf}
\usepackage[colorlinks=true,linkcolor=blue,citecolor=blue]{hyperref}
\usepackage{hyperref}
\usepackage{soul}
\usepackage{ulem}
\usepackage{mathrsfs}
\usepackage{amsmath}

\begin{document}
\title{Series-Expansion Thermal Tensor Network Approach for Quantum Lattice Models}

\author{Bin-Bin Chen}
\affiliation{Department of Physics, Key Laboratory of Micro-Nano Measurement-Manipulation and Physics (Ministry of Education), Beihang University, Beijing 100191, China}

\author{Yun-Jing Liu}
\affiliation{Department of Physics, Key Laboratory of Micro-Nano Measurement-Manipulation and Physics (Ministry of Education), Beihang University, Beijing 100191, China}

\author{Ziyu Chen}
\affiliation{Department of Physics, Key Laboratory of Micro-Nano Measurement-Manipulation and Physics (Ministry of Education), Beihang University, Beijing 100191, China}

\author{Wei Li}
\email{w.li@buaa.edu.cn}
\affiliation{Department of Physics, Key Laboratory of Micro-Nano Measurement-Manipulation and Physics (Ministry of Education), Beihang University, Beijing 100191, China}
\affiliation{International Research Institute of Multidisciplinary Science, Beihang University, Beijing 100191, China}

\begin{abstract}
In this work we propose a series-expansion thermal tensor network (SETTN) approach for efficient simulations of quantum lattice models. This continuous-time SETTN method is based on the numerically exact Taylor series expansion of equilibrium density operator $e^{-\beta H}$ (with $H$ the total Hamiltonian and $\beta$ the imaginary time), and is thus Trotter-error free. We discover, through simulating XXZ spin chain and square-lattice quantum Ising models, that not only the Hamiltonian $H$, but also its powers $H^n$, can be efficiently expressed as matrix product operators, which enables us to calculate with high precision the equilibrium and dynamical properties of quantum lattice models at finite temperatures. Our SETTN method provides an alternative to conventional Trotter-Suzuki renormalization group (RG) approaches, and achieves an unprecedented standard of thermal RG simulations in terms of accuracy and flexibility.
\end{abstract}
\maketitle

\textit{Introduction}.--- Developing accurate methods in quantum manybody computations intrigues people in contemporary condensed matter physics and also other relevant fields, which at the same time also poses a great challenge. In particular, efficient simulations of thermodynamics \cite{TMRG,LTRG,Suzuki-1976,SSE} are indispensable to relate theoretical models to experimental measurements at finite temperatures \cite{TMRGApp,Yan-2012}, and are also of fundamental interest. However, exact diagonalization (ED) calculations of manybody systems encounter the exponentially large Hilbert space, and one needs the full energy spectrum to compute the partition function and other thermodynamic quantities, constituting a severe constraint for large-scale simulations.

Therefore, people resort to various efficient numerical methods to tackle manybody lattice models like spin-1/2 XXZ Heisenberg chain whose Hamiltonian is
\begin{equation}
H = \sum_{\langle i, j \rangle} h_{i,j} = \sum_{\langle i, j \rangle} J (S_i^x S_j^x + S_i^y S_j^y + \Delta S_i^z S_j^z),
\label{Eq:Hamiltonian}
\end{equation}
where $\{S_x, S_y, S_z\}$ are spin operators, $J=1$ is the exchange coupling constant, and $\Delta$ labels the $z$-axis anisotropy. Following Feynman's path integral representation, the partition function (in canonical ensemble) $Z = \mathrm{Tr} (e^{-\beta H})$ can be expressed as a discrete-time thermal tensor network (TTN) by Trotter-Suzuki decomposition $\rho=e^{-\beta H} \simeq (\Pi_{\langle i,j \rangle} e^{-\tau h_{i,j}})^M$, where $\tau$ is a small Trotter slice and $M \tau = \beta$ \cite{Suzuki-1976,Trotter-1959}. In this framework, to calculate thermal quantities such as energy and entropy, people developed stochastic approaches, namely, world-line quantum Monte Carlo (QMC) samplings \cite{Suzuki-1976, Hirsch-1982}, as well as deterministic renormalization-group (RG) methods including transfer-matrix RG (TMRG) \cite{TMRG}, linearized tensor RG (LTRG) \cite{LTRG}, finite-temperature density matrix RG \cite{DMRG, Feiguin-2005}, etc. The RG methods, due to their high accuracy (sampling-error free) and wide versatility (sign-problem free), constitute an important class of methods in computing thermodynamics of low-dimensional quantum lattice models, especially in 1D and quasi-1D.

Nevertheless, these conventional RG approaches, despite their nice accuracies, suffer from two kinds of errors, i.e., Trotter and truncation errors. As a result, compared to ground state simulations, the accuracies of finite-$T$ calculations were dropped behind. The former, either for a finite or infinite 1D system, can reach accuracy of 8 $\sim$ 9 significant digits (even not machine accuracy), with only moderate efforts \cite{DMRG}. However, in discrete-time thermal RG calculations, where the Trotter slice $\tau$ is typically set as $0.05$ or $0.1$ \cite{TMRG,LTRG}, the relative errors are in the order of $10^{-4\sim-5}$ (in free energy). One can choose smaller step $\tau$ to reduce the Totter error and improve the accuracy, but then significantly more RG steps $M$ are demanded correspondingly, making it impractical this way \cite{FootNoteTSErr}. More severely, the two errors may have different signs and the accuracies of thermodynamic properties at any given temperature are not guaranteed to improve by keeping more RG states in the calculations \cite{LTRG}. Therefore, it is appealing if a continuous-time TTN method, which completely gets rid of the Trotter errors, can be devised.

This is possible by noticing that series-expansion QMC realizes Trotter-error-free calculations in discrete forms. The initial idea of series-expansion QMC was proposed by Handscomb in 1960's, who noticed that Taylor expansion can be incorporated in QMC to simulate quantum ferromagnetic spin chains \cite{Handscomb}. Though the utility of Handscomb's method was quite limited, through some efforts \cite{Lyklema-1982, Lee-1984}, it thrives and develops into one of mainstream QMC techniques, dubbed as stochastic series expansion (SSE) \cite{SSE}.

In this work, inspired by the Handscomb's and also SSE methods, we propose a series-expansion tensor network method for quantum manybody systems. Compared to previous Trotter-Suzuki RG methods, by exploiting an efficient tensor-network representation of the powers of total Hamiltonian ($H^n$), we remove the Trotter errors completely and achieve an unprecedentedly high accuracy in computing thermodynamic properties. Compared to SSE, our method employs a deterministic RG approach to do the series-expansion calculations, which is in principle sign-problem free for frustrated spin models. Moreover, time-dependent correlation functions are also accessible in this method, again calculated in a Trotter-error-free manner.

\textit{Series-expansion thermal tensor networks}.--- Utilizing a Taylor (precisely Maclaurin) series expansion, the partition function can be written as:
\begin{equation}
Z(\beta) \simeq \sum_{n=0}^{N} \kappa_{n}(\beta) = \sum_{n=0}^N \omega_n(\beta) \mathrm{Tr}(H^n),
\label{Eq:SE}
\end{equation}
where $\omega_n(\beta)=(-\beta)^n/n!$, $n$ runs over the expansion index space and $N$ is the cut-off order of $n$. Owing to the fact that $\omega_n(\beta)$ obeys Poisson distribution and centers at $\langle n \rangle = \beta$, when $N$ is sufficiently large the expansion is essentially free of cut-off error. It thus offers a continuous-time approach (with $\beta$ regarded as imaginary time) through a discrete formalism. In Handscomb's method (and also in SSE), the total $H$ is decomposed into a sum of local terms, and the QMC samplings are performed in the space of index sequences \cite{Handscomb, SSE}.

\begin{figure}[tbp]
  \includegraphics[angle=0,width=0.95\linewidth]{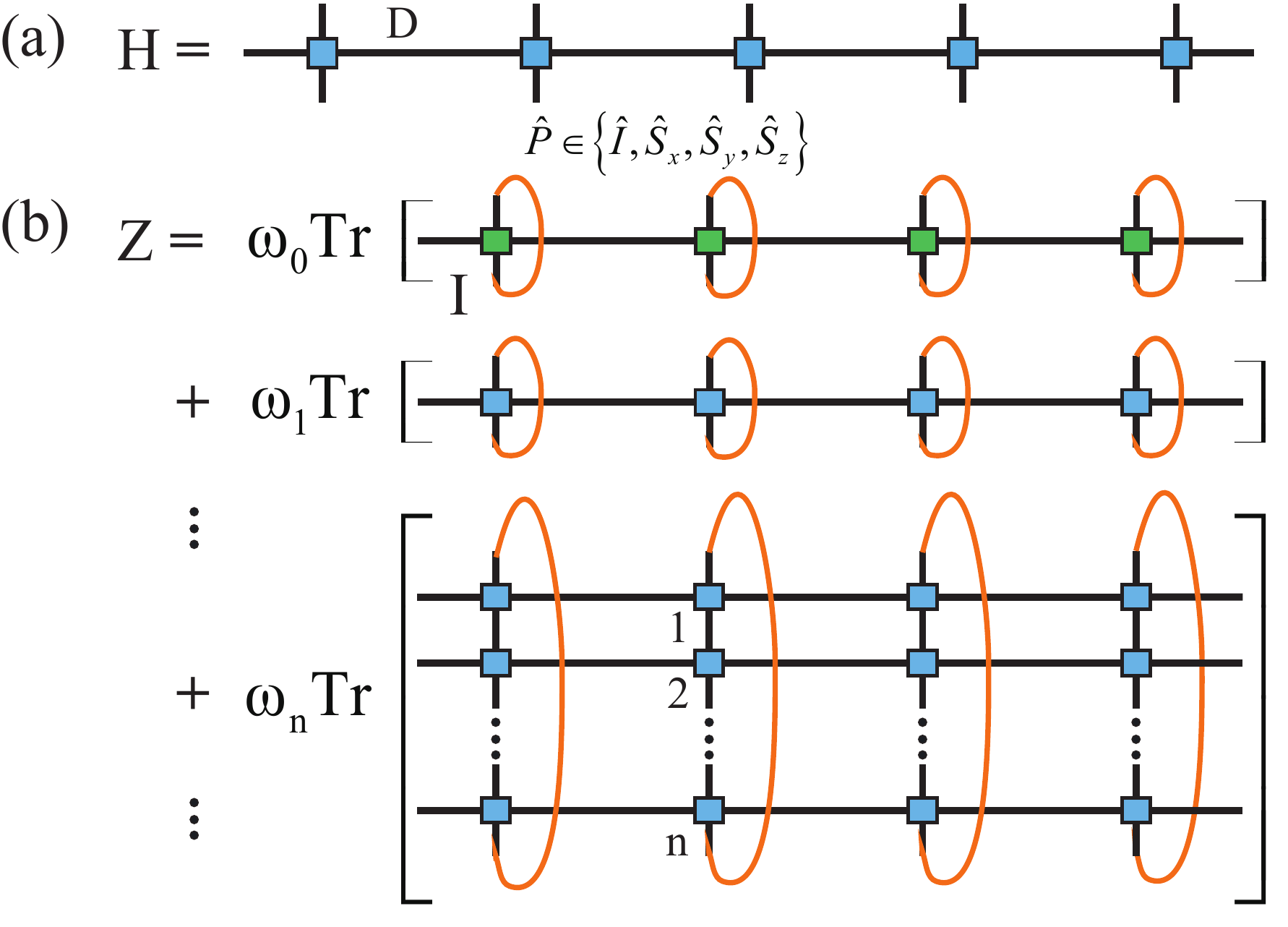}
  \caption{(Color online) (a) MPO representation of total Hamiltonian of Heisenberg spin chain model, $D$ labels geometric bond dimension, and local tensor $P$ comprises spin operators and an identity. (b) The partition function $Z$ can be expressed as a sum of $\omega_n \mathrm{Tr} (H^n)$ terms, each of which constitutes a tensor network.}
  \label{Fig:SETTN}
\end{figure}

We point out that the series expansion Eq. (\ref{Eq:SE}) can be actually encoded in an efficient representation without splitting $H$ into local terms, in the framework of tensor networks (TNs). Exploiting this fact, we propose a continuous-time algorithm named series-expansion TTN (SETTN).  Specifically, for 1D models like XXZ spin chain, one can construct a matrix product operator (MPO) with small bond dimensions to represent the total Hamiltonian Eq. (\ref{Eq:Hamiltonian}). As shown in Fig. \ref{Fig:SETTN}(a), the local $P$ tensor stores the spin and identity operators $\{\hat{S}_x, \hat{S}_y, \hat{S}_z, \hat{I}\}$, which are all $d \times d$ matrices ($d$ the dimension of local Hilbert space). $P$ tensors are connected via geometric bonds to constitute a MPO, with bond dimension $D=5$(4) for XXZ(XY) model. For instance, in XY-chain model under open boundary conditions (OBCs) the nonzero elements in $P$ tensor are $P_{1,1}=P_{4,4}=\hat{I}, P_{1,2} = P_{2,4} = \hat{S}_x, P_{1,3}=P_{3,4}=\hat{S}_y$, and all operators in $P$ apply vertically as shown in Fig. \ref{Fig:SETTN}(a).

Despite some studies on the MPO representation of $H$, in the context of ground state and also time evolution problems \cite{Frowis-2010,Pirvu-2010,Zaletel-2015}, there are still much to be explored regarding the MPO representation of its power $H^n$. In Fig. \ref{Fig:SETTN}(b), we plot the $H^n$ series, each term of which can be represented as a TN. The partition function can be obtained by summing over the trace of each TN [multiplied by $\omega_n(\beta)$ in Eq.(\ref{Eq:SE}], given that we can perform efficient contractions of them. This constitutes our main idea of SETTN method.

\begin{figure}[tbp]
  \includegraphics[angle=0,width=1\linewidth]{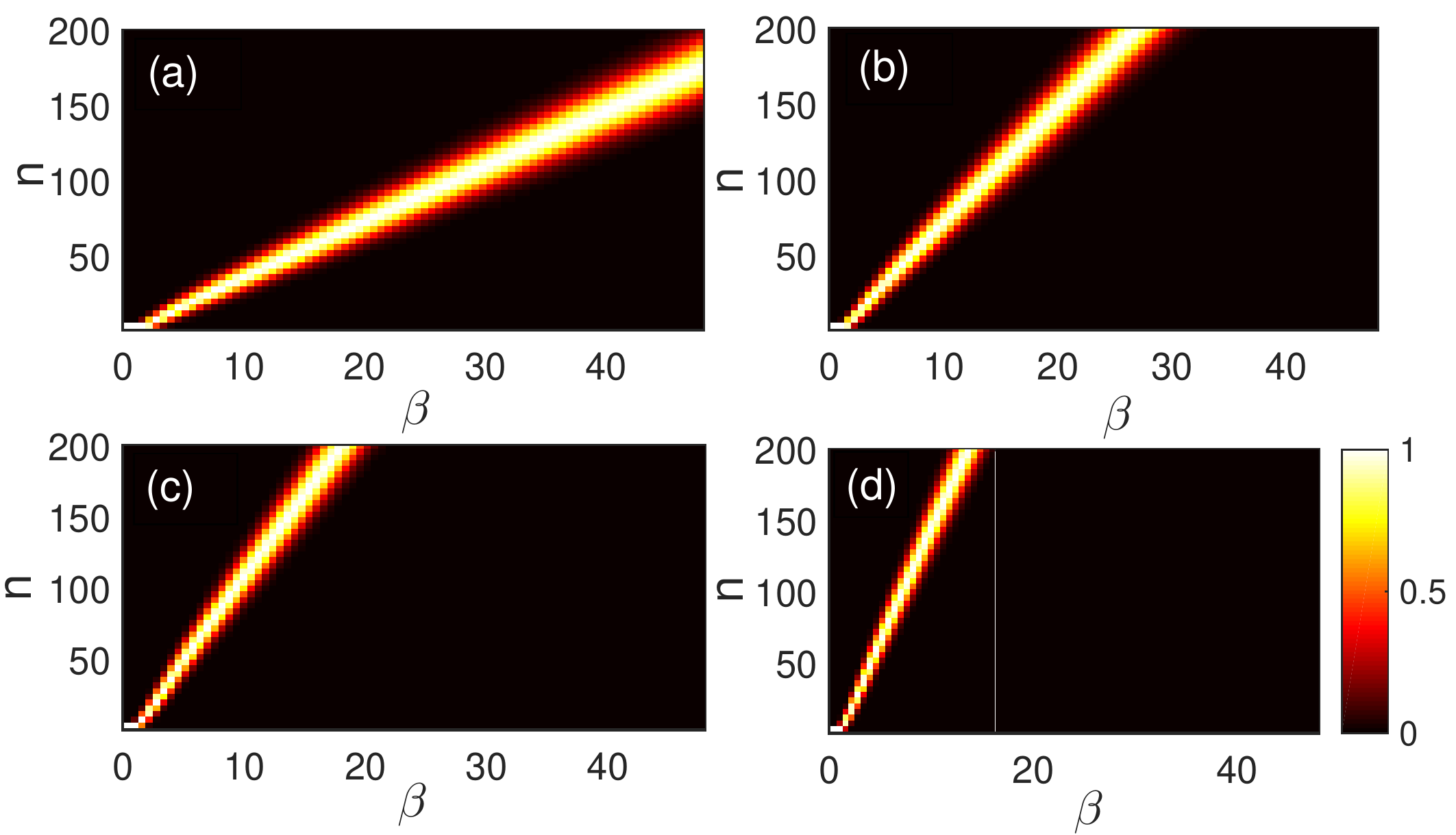}
  \caption{(Color online) Normalized weights $\kappa_n(\beta)$ (for each fixed $\beta$, the $\kappa_n(\beta)$ series is normalized such that the maximal value equals unity) in partition function Eq. (\ref{Eq:SE}) for various system sizes $L=12,24,36,48$, from (a) to (d), respectively. The sharp contrast of brightness (intensities of weights) reveals that only a relatively small number of $n$ terms are significant, around the linear ``light" cones whose slops are 3.55, 7.69, 11.17, 14.61, from (a) to (d), respectively.}
  \label{Fig:Wght}
\end{figure}

Two concerns immediately arise: Firstly, in order to get a converged and accurate summation for $Z$, one may wonder how many terms ($N$) are generically sufficient, and how does $N$ scale with $\beta$ and system size $L$; The second possible worry is about the compressibility of $H^n$ series. To answer the first question, in Fig. \ref{Fig:Wght} we show the properly normalized weights $\kappa_n(\beta)$ in Eq. (\ref{Eq:SE}) for various system sizes. It can be seen that, for every $\beta$, only a small number of weights (in the bright cone-like region) essentially contribute to $Z$. The average $\langle n \rangle_{H}$, where the cone center resides, increases linearly with $\beta$. In addition to that, the slop of the cone increases linearly with system size $L$ as well, and we can estimate that $\langle n \rangle_{H} \approx \beta L/3$ from Fig. \ref{Fig:Wght}. This is remarkable since it means that in order to obtain thermal properties of an $L$-site system at inverse temperature $\beta$, one only needs to prepare $O(\beta L)$ number of $H^n$ terms. We also numerically calculated XXZ model with $\Delta$ other than 0, and find this relation always valid.

Actually, such an observation can be understood as following: $\mathrm{Tr} (H^n)$ is proportional to $|E_{ln}|^n$, where the extensive quantity $E_{ln} = e_{ln} L$ is the energy eigenvalue of $H$ with largest norm. Therefore, the dominant term in series $\kappa_n(\beta)= \frac{(\beta L |e_{ln}|)^n}{n!}$ should be at $\langle n \rangle_H = \beta L |e_{ln}|$. Remember that the ground-state energy per site (in the thermodynamic limit) of XY-chain is $-1/\pi$ (which happens to be also the eigenvalue with largest norm), thus $\langle n \rangle_H \simeq \beta L/\pi$, in perfect agreement with the numerical results in Fig. \ref{Fig:Wght}.

\begin{figure}[tbp]
  \includegraphics[angle=0,width=1\linewidth]{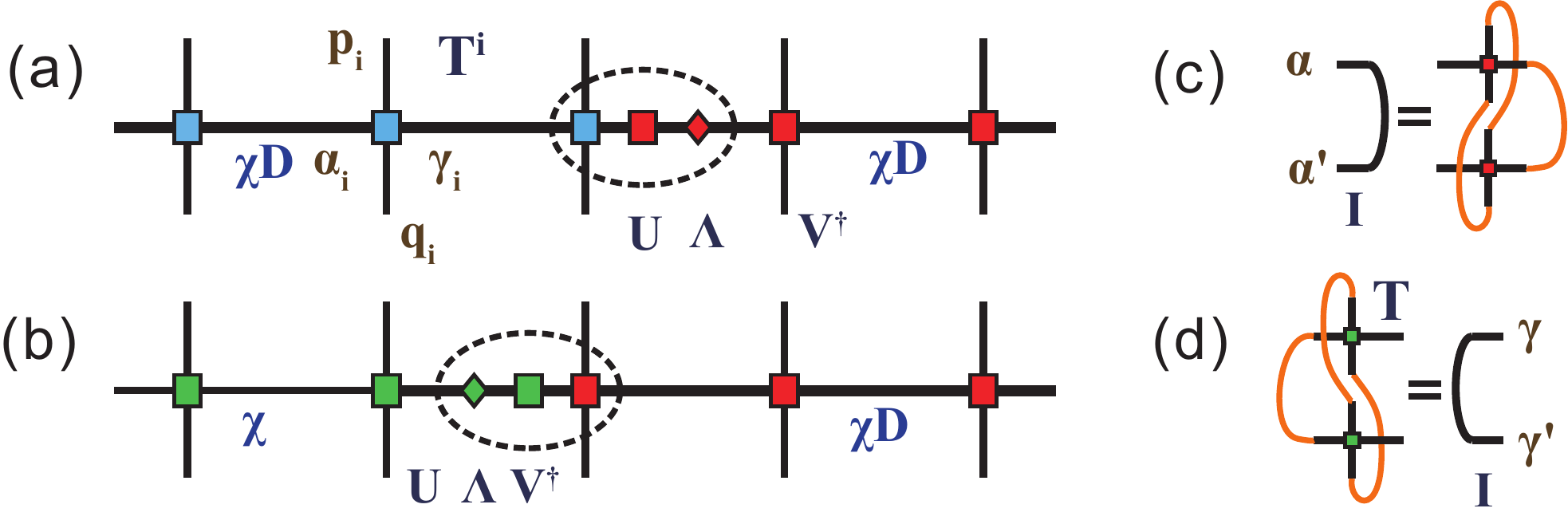}
  \caption{(Color online) (a) Left and (b) right sweep, bringing the enlarged MPO into canonical form and perform optimal truncations. Diagrammatic representation of (c) right- and (d) left-canonical condition, $I$ is identity matrix.}
  \label{Fig:Canon}
\end{figure}

\begin{figure}[tbp]
  \includegraphics[angle=0,width=1\linewidth]{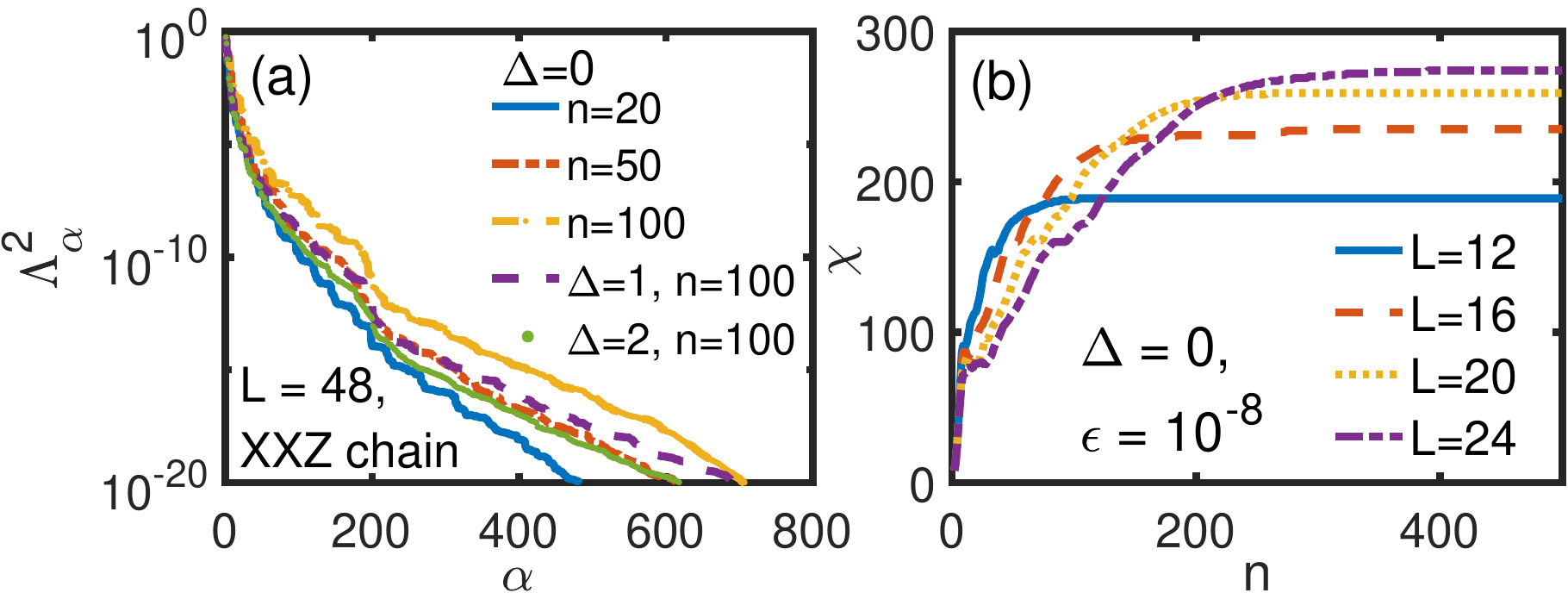}
  \caption{(Color online) (a) Fast decaying ``entanglement" spectrum of MPO $H^n$ ($n=20,50,100$), cut in the middle of the $L=48$ XXZ chains with $\Delta=0$ (XY), 1 (Heisenberg), and 2 (XXZ). (b) shows the required MPO bond dimensions $\chi$ to maintain a fixed truncation error $\epsilon=10^{-8}$ in XY chain of $L=12,18,20,24$.}
  \label{Fig:Comp}
\end{figure}

\textit{Canonical optimization and compressibility of MPO $H^n$}.--- Now we address the second question on whether each $H^n$ has an efficient MPO representation (compressibility) and how the bond dimensions $\chi$ scale as order $n$, as well as system size $L$, increases.

As afore mentioned, $H$ has a simple MPO representation, which serves as a good starting point to prepare the $H^n$ series [in Fig.\ref{Fig:SETTN}(b)]. Suppose we have obtained MPO of $H^{n-1}$ (with bond dimension $\chi$), in order to get the next MPO $H^n$ we have to project a single $H$ onto $H^{n-1}$, obtaining an MPO with ``fat"  bond dimension $\chi D$. In order to prevent bond dimensions from this exponential growth as $n$ increases, a proper truncation here is crucial.

Note that the enlarged MPO $H^n$ generically deviates from the so-called canonical form, and thus we first perform a right-to-left sweep process, gauging the tensors into their right-canonical form. In each single operation, we reshape $T^i_{\alpha_i, p_i, q_i, \gamma_i}$ into a matrix and take a singular value decomposition $T^i_{\alpha_i; (p_i,q_i,\gamma_i)} = \sum_{\tilde{\alpha}_i=1}^{\chi D} U_{\alpha_i, \tilde{\alpha}_i} \Lambda_{\tilde{\alpha}_i} V^{\dag}_{\tilde{\alpha}_i, (p_i,q_i,\gamma_i)}$. Then, we update $\tilde{T}^i=V^{\dag}$, and associate $U_{\alpha_i, \tilde{\alpha}_i}  \Lambda_{\tilde{\alpha}_i}$ to the $T^{i-1}$ on the left [as shown in Fig. \ref{Fig:Canon}(a)]. It is clear that $\tilde{T}^i$ has been gauged into a right canonical form, i.e., $\sum_{(p_i,q_i,\gamma_i)} \tilde{T}^i_{\alpha, (p_i,q_i,\gamma_i)} \tilde{T}^{i *}_{\alpha', (p_i,q_i,\gamma_i)} = \delta_{\alpha, \alpha'}$ [Fig. \ref{Fig:Canon}(c)]. We iteratively perform this canonicalization process on each site, all the way from right to left, gauging all tensors to their right canonical forms. Subsequently, we perform a truncation sweep, analogous to the previous procedure, but from left to right. As this right-moving sweep proceeds, we have all left-canonical tensors [see Fig. \ref{Fig:Canon}(d)] to the left of $\Lambda_{\alpha_i}$ and all tensors right-canonical on the right-hand side. Note that under this condition SVD actually plays the role as a global Schmidt decomposition of the supervector $H^n$, and the singular values in $\Lambda$ are the (nonnegative) Schimdt weights, providing a quasi-optimal truncation criteria. We keep the important bond bases corresponding to large $\Lambda_{\alpha}$ and discard those having small weights.

In Fig. \ref{Fig:Comp}(a) we plot the normalized $\Lambda^2_{\alpha}$ in obtained canonical MPOs $H^n$, from which we see very fast decaying weights. Therefore, it is only necessary to keep a small number ($\chi$) of states out of the exponentially large bond space, while obtaining a good approximation to full $H^n$. In Fig. \ref{Fig:Comp}(b) we show that to maintain the fixed truncation error $\epsilon$ (sum of discarded $\Lambda_{\alpha}^2$ weights, $\epsilon=10^{-8}$ here), the required bond dimension $\chi$ firstly increases and then saturates as order $n$ increases. It is also observed that as the system size $L$ enhances, both the saturation $\chi$ and convergence point $n$ increases, revealing that the convergence behavior of $\chi$ is actually due to finite-size effects. Interestingly, for small n (say, $<100$), required $\chi$ decreases as $L$ enlarges; while for very large $n$ one needs larger $\chi$ to maintain the same truncation error for longer $L$. Overall, in the models examined, we find that the MPO indeed constitutes an efficient representation for $H^n$, for any $n$ and $L$ \cite{Footnote2norm}.

In conventional Trotter-Suzuki TTN methods (TMRG, LTRG, etc), time costs of a local operation scale as $O(\chi^3 d^6)$; In the SETTN algorithm, this time cost scales in the same order on $\chi$, as $O(\chi^3 D^3 d^2)$. The memory cost of SETTN [$O(\chi^2 D^2 d^2)$] is also slightly larger than that of TMRG/LTRG [$O(\chi^2 d^4)$]. The number of total projection steps in TMRG/LTRG calculations is $M=\beta/\tau$. For instance, when largest $\beta=50$ and $\tau=0.05$, then $M=1000$. In SETTN calculations, take XY-chain as an example, total number of projection steps (to obtain $H^n$) is $\sim \beta L/3$ (Fig. \ref{Fig:Wght}), which is less(larger) than 1000 given $L<60(>60)$. To conclude, SETTN bears marginally larger time and memory costs than LTRG, while achieves the flexibility of continuous (imaginary) time.

\begin{figure}[tbp]
  \includegraphics[angle=0,width=1\linewidth]{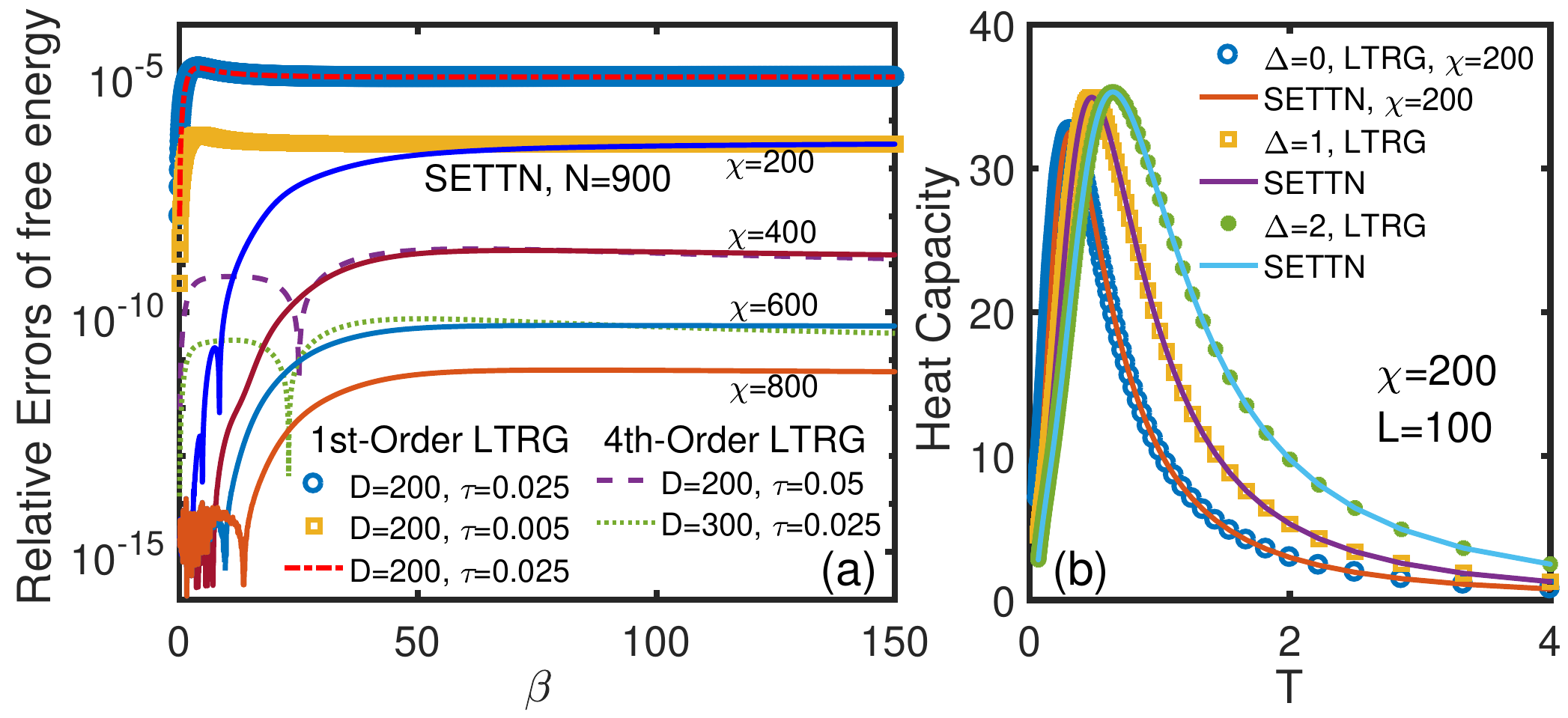}
  \caption{(Color online) (a) Relative errors of free energy of $L=14$ XY chain, obtained by SETTN and LTRG, down to ultra-low temperatures. (b) Heat capacity of $L=100$ XXZ chains of $\Delta=0,1,2$, where the SETTN and LTRG results agree perfectly with each other.}
  \label{Fig:Rslt}
\end{figure}

\textit{Thermodynamics results}.--- In Fig. \ref{Fig:Rslt}(a), compared to exact diagonalization (ED) data $F_{\mathrm{ED}}$, we include errors of free energy $F$, $|F-F_{\mathrm{ED}}|/|F_{\mathrm{ED}}|$, calculated by SETTN and LTRG with both the first- and fourth-order Trotter-Suzuki decompositions on a XY chain ($\Delta=0$). We find that the accuracy of SETTN improves continuously as $\chi$ enhances, and the relative error reaches $10^{-11}$ when $\chi=800$ even at ultra-low temperatures ($\beta=150, T/J\simeq0.0067$), which is $5\sim6$ orders of magnitude smaller than that of LTRG using first-order decomposition \cite{Xiang-2016}. The latter suffers from the Trotter error, which is $\sim 10^{-5}$ for $\tau=0.025$ (and $10^{-6}$ for $\tau=0.005$), and cannot be improved by increasing $\chi$. Note that $\tau=0.005$ is even not very practical in real-life Trotter-Suzuki RG calculations since it leads to too many projection steps (thus costs 8 times the CPU time SETTN consumes with $\chi=200$). To reach similar accuracy as that of SETTN with $\chi > 200$, extremely small Trotter slice $\tau$ is required in the first-order decomposition, making it very inefficient and impractical. We have also compared the fourth-order decomposition with SETTN in the high accuracy regime. To achieve the accuracy of $\sim 10^{-9}$ ($10^{-10}$), the fourth-order calculation takes 4(5) times the CPU time SETTN costs. Therefore, the SETTN approach proposed in this work, we think, establishes an unprecedented standard of finite-temperature calculations, surpassing previous (world-line) TTN methods in both accuracy and efficiency.

Thermodynamic properties other than free energy can also be conveniently evaluated in the SETTN framework. For instance, the energy $E(\beta) = \frac{\mathrm{Tr} (H e^{-\beta H})}{ \mathrm{Tr} (e^{-\beta H})} = \frac{\sum_{n=0}^N \omega_n(\beta) \mathrm{Tr} (H^{n+1})}{\sum_{n=0}^N \omega_n(\beta) \mathrm{Tr} (H^n)}$
can be easily calculated with obtained $\mathrm{Tr} (H^n)$ series. In Fig. \ref{Fig:Rslt}(b) we show results of heat capacity $C = -\beta^2 (\frac{\partial E}{\partial \beta})$ of large-scale ($L=100$) XXZ chain with $\Delta=0,1,2$. Since ED is no longer possible for such large system, we compare SETTN and LTRG (adapted to finite-size systems) results, and see perfect agreements in all cases.

\begin{figure}[tbp]
  \centering
  \includegraphics[angle=0,width=1.0\linewidth]{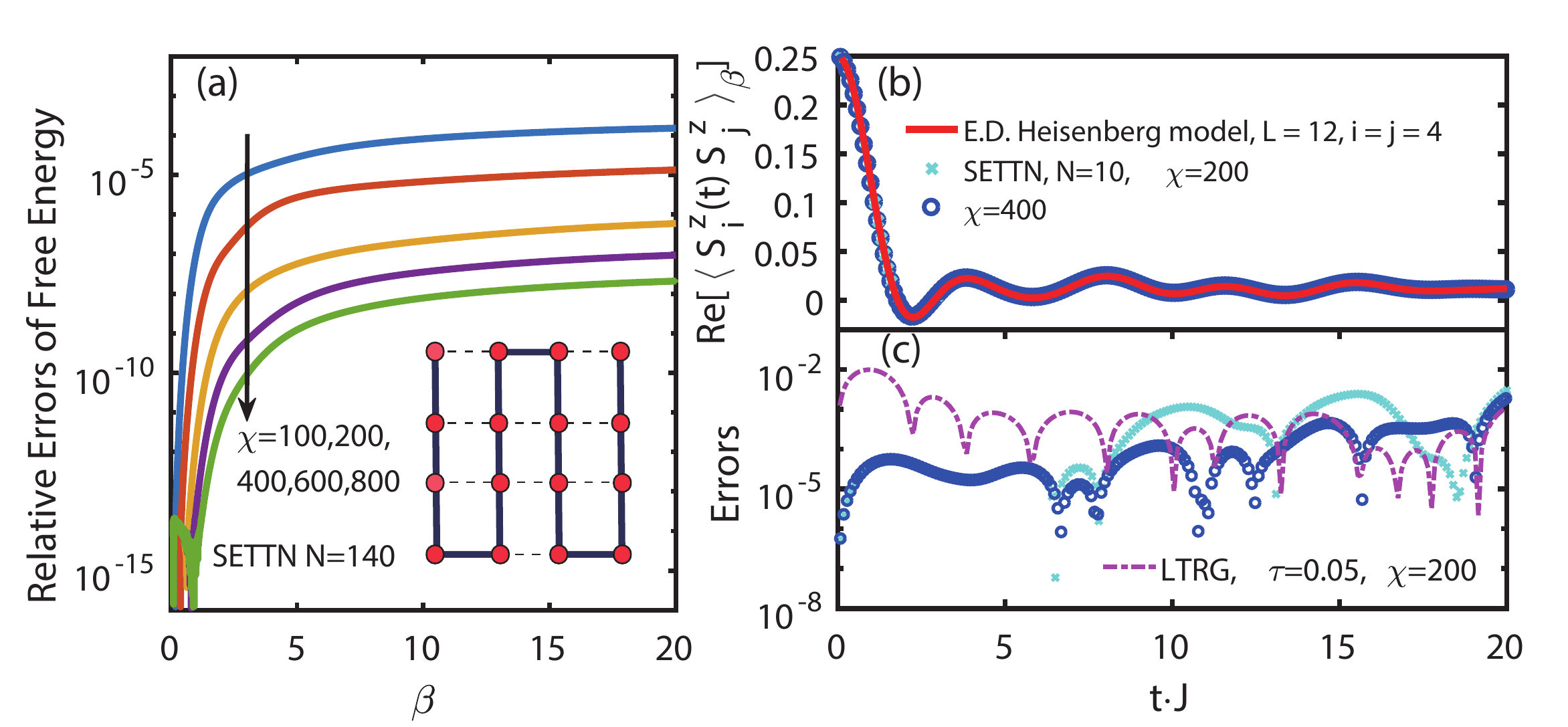}
  \caption{(Color online) (a) The relative errors of free energy in SETTN calculations, with $\chi=100,200,400,600,800$, respectively, for a 4$\times$4 square-lattice Ising model $H = \sum_{\langle i,j \rangle} S_i^z S_j^z -h \sum_i S_i^x$ (with $h=1$). (b) The real part of time-dependent correlation functions $\langle S_i^z(t) S_j^z(0) \rangle_{\beta}$ at $\beta=1$. (c) The errors of SETTN calculations with $\chi=200, 400$, respectively, compared to the LTRG ones with time step $\delta=0.05$.}
  \label{Fig:DynamCF}
\end{figure}

Not limited in 1D, SETTN method can also be generalized to calculate lattice models of higher dimensions  \cite{FootNote2}. In Fig. \ref{Fig:DynamCF}(a) we show the results of the quantum Ising model on a $4 \times 4$ square lattice, whose Hamiltonian can be represented by MPO following a snake-like line, as depicted in the inset. As in 1D, the relative errors of free energy results (compared to ED data) improve continuously as $\chi$ enhances, and are stimulatingly small for even moderate $\chi$'s, revealing that SETTN can be employed to calculate thermodynamics of 2D lattice models with high accuracy.

\textit{Time-dependent correlation functions}.--- With efficient MPO representations of $H^n$ series at hand, we can recycle them to calculate dynamical correlation functions $\langle S_i^z(t) S_j^z(0) \rangle_{\beta} = \mathrm{Tr}(e^{-\beta H} e^{i H t} S_i^z e^{-i H t} S_j^z)$. By expanding $e^{iH\delta}=\sum_{n=0}^N \frac{(i\delta)^n}{n!} H^n$ at a relatively small time $\delta$ ($=0.1$, say, in practical simulations), we can find a Trotter-error-free MPO representation $Q(\delta)$ for $e^{iH\delta}$, by minimize the distance of $Q(\delta)$ with the $H^n$ series. Given an accurate $Q(\delta)$ obtained, the Heisenberg operator $\tilde{O}(-t) = e^{-iHt}\hat O e^{iHt} = [Q(\delta)^{\dagger}]^K \hat O [Q(\delta)]^K$ can also be efficiently expressed as an MPO [where $K=t/\delta$, $\hat O=\rho, S_i^z$, etc.], with which the dynamical correlation function can be evaluated as $\mathrm{Tr}\{\tilde{\rho}(\beta,-t) S_i^z \tilde{S}_j^z(-t)\}$.

Figs. \ref{Fig:DynamCF}(b,c) show the real part of $\langle S_i^z(t) S_j^z(0) \rangle_{\beta}$ for a Heisenberg chain model at finite temperature ($\beta=1$). In the calculations, we expand the exponential operators $Q(\delta)$ and $\rho(\beta)$ up to $N=10$ orders, and retain $\chi=200, 400$ states on the bonds of the MPOs $\tilde{O}(-t)$ ($O=\rho,S^z_j$). Fig. \ref{Fig:DynamCF}(b) shows that, for time up to $tJ=20$, the results agree perfectly with the ED data. In Fig. \ref{Fig:DynamCF}(c), comparing with Trotter-Suzuki calculations (LTRG plus real-time evolution with step $\delta=0.05$) \cite{Karrasch-2012, Banuls-2009}, we find the accuracies of SETTN results are better than LTRG in short-time regime and of comparable accuracies for relatively long time.


\textit{Conclusion and outlook}.--- We propose a series-expansion thermal tensor network (SETTN) approach for accurate calculations of equilibrium and dynamical properties of quantum manybody systems. We show that (a) only a small number of weights in the summation $\sum_{n=0}^N \frac{(-\beta)^n}{n!} \mathrm{Tr} (H^n)$ contribute to the partition function, and the upper bound of order is $N \sim \beta L |e_{ln}| $, where $e_{ln}$ is the energy eigenvalue of largest norm; and (b) each $H^n$ (regardless of the length $L$) can be efficiently expressed as a matrix product operator, judged from its fast-decaying ``entanglement" spectrum. With these two nice properties, we are able to realize a Trotter-error-free continuous-time thermal tensor network method, whose extraordinarily high accuracy and flexibility are benchmarked by calculations of XXZ spin chains and square-lattice quantum Ising model.

There are still some open questions deserving further investigations. To name a few, application of SETTN method to more challenging 2D systems is surely appealing; Implementation of symmetries can improve the efficiency in SETTN; In the Trotter-Suzuki framework, evaluating thermodynamic properties of models with long-range interactions is tricky, whereas since the MPO representations of such Hamiltonians can be systematically constructed \cite{Pirvu-2010}, applying SETTN approach to such systems is inspiring; Besides Taylor expansion, exploiting instead Chebyshev expansion in SETTN is also quite motivated \cite{vDelft, Braun-2014, Tiegel-2014, Bruognolo-2016}.

\begin{acknowledgments}
\textit{Acknowledgments}.--- WL is indebted to Andreas Weichselbaum, Jan von Delft, Yong-Liang Dong, Hong-Hao Tu, Lei Wang, and Zi-Yang Meng for stimulating discussions. This work was supported by the National Natural Science Foundation of China (Grant Nos. 11504014, 11274033, 11474015 and 61227902), the Research Fund for the Doctoral Program of Higher Education of China (Grant No. 20131102130005), and the Beijing Key Discipline Foundation of Condensed Matter Physics.
\end{acknowledgments}


\begin{thebibliography}{99}
\bibitem{Footnote2norm} Note that the compression scheme is based on the Frobenius norm of MPO, which is convenient to deal with in the tensor network algorithms and are widely adopted to compress MPOs. Nevertheless, an optimization scheme using instead the operator norm still remains to be explored.
\bibitem{FootNote2} For 2D system, we can make use of a MPO or tensor product operator representation for the Hamiltonian: in the former, the system is treated as 1D chain with ``long-range" interactions, and thus the bond dimension of MPO scales linearly as the width of the lattice; while the latter has a fixed bond dimension regardless of the system size. The details of 2D SETTN calculations will be published elsewhere.
\end{thebibliography}
\end{document}